\documentclass[aip, apl, reprint, amsmath,amssymb,superscriptaddress]{revtex4-1}
\usepackage{graphicx}
\usepackage[colorlinks=true,citecolor=blue]{hyperref}
\usepackage[normalem]{ulem}
\usepackage{xcolor}
\usepackage{comment}
\usepackage{cancel}
\newcommand*{\citen}[1]{%
  \begingroup
    \romannumeral-`\x % remove space at the beginning of \setcitestyle
    \setcitestyle{numbers}%
    \cite{#1}%
  \endgroup   
}

\newcommand{\tNb}{$t_\mathrm{Nb}$}
\newcommand{\Bres}{$B_\mathrm{res}$}
\newcommand{\LW}{$\Delta B$}

\draft % marks overfull lines with a black rule on the right

\begin{document}

\title{Absence of detectable spin and orbital pumping from Ni to Nb by out-of-plane ferromagnetic resonance}

\author{Omolara A. Bakare}
\email{obakare@vt.edu}
\author{Galen T. Street}
\author{Sachli Abdizadeh}
\author{Rachel E. Maizel}
\altaffiliation{}
\affiliation{Department of Physics, Virginia Tech, Blacksburg, VA, USA}
\author{Christoph Klewe}
\affiliation{Advanced Light Source, Lawrence Berkeley National Laboratory, Berkeley, CA, USA}
\author{Satoru Emori}
\email{semori@vt.edu}
\affiliation{Department of Physics, Virginia Tech, Blacksburg, VA, USA}
\affiliation{Academy of Integrated Science, Virginia Tech, Blacksburg, VA, USA}
%\thanks{}

\date{\today}

\begin{abstract}
Excited ferromagnets can pump spin angular momentum, along with possibly orbital angular momentum. Among elemental ferromagnets, Ni has been proposed to exhibit substantial orbital pumping relative to spin pumping. 
Here, we search for a signature of orbital pumping by Ni, specifically by comparing out-of-plane ferromagnetic resonance in heterostructures without Ni (FeV/Nb) and with Ni (FeV-Ni/Nb). 
The FeV/Nb series shows a clear increase in Gilbert damping with the Nb sink thickness, attributed to spin pumping from FeV to Nb. Surprisingly, the FeV-Ni/Nb series exhibits no such damping increase, revealing no significant spin or orbital pumping from Ni to Nb. 
Our results offer a fresh perspective on angular-momentum transfer in Ni-based heterostructures, suggesting that the interpretation of some strong orbitronic effects may require further consideration.
\end{abstract}

\maketitle

%%% Indroduction %%%
A flow of angular momentum can exert a torque on the magnetization, thereby controlling the digital states of nanomagnetic memories~\cite{brataas2012current, manchon2019current}. In addition to the flow of spin angular momentum, recent studies suggest that the flow of orbital angular momentum may play a critical role~\cite{go2021orbitronics, wang2024orbitronics,burgos2024orbital}. %In principle, orbital current may transport more than $\hbar/2$ of angular momentum per carrier, exceeding the fundamental limit of spin current. 
For example, torques due to orbital Hall effects in some metals, such as Ti and Nb, are reportedly an order of magnitude stronger than those due to spin Hall effects~\cite{dutta2022observation,bose2023detection,liu2023giant,hayashi2023observation,choi2023observation}. 
However, distinguishing spin- and orbital-current effects is fundamentally a challenge~\cite{lee2021orbital, sala2022giant, han2025orbital}, hindering the design of spin/orbitronic devices. Coexisting charge-spin conversion processes in nanomagnetic metal hetrostructures~\cite{davidson2020perspectives,go2020theory,kim2024spin,lee2021orbital,sala2022giant} further complicate the interpretation of spin/orbital torque experiments. 

\begin{figure}[tb]
\centering
  \includegraphics[width=1.0\columnwidth]{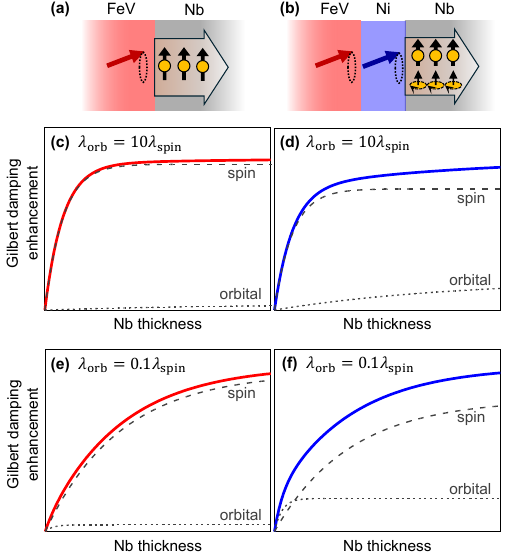}
  \caption{(a,d) Schematics of heterostructures with different sources of pumping: (a) FeV hypothesized with predominantly spin pumping; (b) Ni (coherently coupled to FeV) hypothesized with sizable orbital pumping relative to spin pumping. (c-f) \emph{Hypothesized} schematics of Gilbert damping enhancement vs Nb sink thickness where: (c,d) the orbital diffusion length $\lambda_\mathrm{orb}$ is an order of magnitude greater than the spin diffusion length $\lambda_\mathrm{spin}$ with the (c) FeV source and (d) Ni source;  (e,f) $\lambda_\mathrm{orb}$ is an order of magnitude less than $\lambda_\mathrm{spin}$ with the (e) FeV source and (f) Ni source.}\label{fig:cartoons}
\end{figure}

An alternative approach is to examine spin/orbital \emph{pumping}, which does not introduce any net charge current in the studied heterostructure~\cite{go2025orbital,pezo2025adiabatic,tserkovnyak2005nonlocal,boone2013spin,boone2015spin,santos2023inverse,ding2024mitigation,hayashi2024observation,wang2025orbital}. 
A precessing magnetization in the source material, consisting of spin and orbital moments~\cite{emori2024quantifying, ding2024mitigation}, produces nonequilibrium spin and orbital accumulations at the interface. These accumulations propagate as spin and orbital currents [Fig.~\ref{fig:cartoons}(a,b)] and ultimately decay in the adjacent sink material. The decays of spin and orbital currents result in an additional loss of nonequilibrium angular momentum from the ferromagnetic source, manifesting as an enhancement of Gilbert damping~\cite{tserkovnyak2005nonlocal,haney2013current,boone2013spin,boone2015spin}. Thus, the spin or orbital decay (diffusion) length can be deduced from Gilbert damping vs sink thickness~\cite{tserkovnyak2005nonlocal,haney2013current,boone2013spin,boone2015spin}: the damping first increases with increasing sink thickness as more spin or orbital current decays, and then saturates beyond the diffusion length. 

Here, we test the hypothesis, motivated by the theoretical predictions of Go \textit{et al.}~\cite{go2025orbital}, that the ferromagnetic source dictates the nature of spin/orbital pumping. According to Ref.~\citen{go2025orbital}, the magnitude of orbital pumping may be only a few percent relative to spin pumping for Fe, but it may exceed 10\% for Ni. Hence, we might expect the Gilbert damping enhancement as a function of sink thickness to differ between an Fe-based source [Fig.~\ref{fig:cartoons}(a,c,e)] and a Ni-based source [Fig.~\ref{fig:cartoons}(b,d,f)]. Assuming substantial orbital pumping from Ni, if the orbital decay length is much greater than the spin decay length~\cite{hayashi2023observation,choi2023observation,lyalin2023magneto}, we might observe a gradual increase in Gilbert damping due to orbital pumping over a wide sink thickness range, on top of a sharper increase due to spin pumping [Fig.~\ref{fig:cartoons}(d)]. On the other hand, if the orbital decay length is much shorter than the spin decay length~\cite{belashchenko2023breakdown,rang2024orbital,urazhdin2023symmetry,ning2025orbital}, the observed damping might first increase sharply at small sink thicknesses due to orbital pumping, followed by a more gradual increase due to spin pumping [Fig.~\ref{fig:cartoons}(f)].

We compare the Gilbert damping of two types of heterostructure, as illustrated in Fig.~\ref{fig:cartoons}: 
\begin{enumerate}
\item an Fe$_{70}$V$_{30}$ source interfaced with a Nb sink (FeV/Nb);
\item a bilayer source of Fe$_{70}$V$_{30}$ and Ni, with Ni interfaced with a Nb sink (FeV-Ni/Nb). 
\end{enumerate}
Instead of elemental Fe, we chose Fe$_{70}$V$_{30}$ with a reduced saturation magnetization~\cite{smith2020magnetic,arora2021magnetic} to facilitate reliable quantification of spin/orbital pumping with out-of-plane ferromagnetic resonance (FMR)~\cite{boone2015spin,boone2013spin}. 
A single-layer source of Ni with high intrinsic Gilbert damping (damping parameter $\sim$0.03) would obscure small damping enhancements from pumping~\cite{schoen2017magnetic}. As such, we instead chose FeV-Ni, in which thin Ni is exchange-coupled to low-damping FeV, ensuring a large FMR signal-to-noise ratio~\cite{smith2020magnetic,arora2021magnetic} and precise quantification of damping enhancements from the Ni/Nb interface. 
Nb has attracted much attention for its potentially large orbital Hall effect \cite{dutta2022observation, liu2023giant,keller2025identification}, such that it is an interesting prototypical metal for investigating spin/orbital transport. Our study focuses on Nb as the sink instead of Ti (another interesting metal in orbitronics~\cite{hayashi2023observation,choi2023observation}) because Nb has a lower reactivity with residual gases in the deposition chamber. 

In this letter, we demonstrate that the observed behavior for FeV/Nb is consistent with the hypothesized predominance of spin pumping. Surprisingly, our observation for FeV-Ni/Nb completely deviates from our hypothesis [Fig.~\ref{fig:cartoons}(b,d,f)]; we find no resolvable signature of spin or orbital pumping in FeV-Ni/Nb, as the Gilbert damping enhancement remains essentially zero. Our findings imply minimal angular momentum transport between Ni and Nb, e.g., an order of magnitude smaller than between FeV and Nb,  pointing to the need to re-examine reports of robust orbital transport in Ni/Nb and similar metal heterostructures. 

%%% Sample Growth and Stack Structures%%%
All samples were deposited by magnetron sputtering on Si substrates with 50-nm-thick thermally grown SiO$_2$, unless otherwise noted. The base pressure before deposition was $\lesssim 8\times10^{-8}$ Torr. The Ar sputtering gas pressure was 3 mTorr. The FeV/Nb series has the stack structure of substrate / Ti(3) / Cu(3) / Fe$_{70}$V$_{30}$(20) / Nb(\tNb) / TiO$_x$(3), where the values in parentheses denote thicknesses in nm. The Nb thickness \tNb\ varies from 0 to 40 nm. 
The Ti/Cu seed layer enables smooth textured growth of the subsequent polycrystalline film layers, yielding narrower FMR linewidths (higher signal-to-noise ratios)~\cite{edwards2019co}.  The TiO$_x$ capping layer (naturally oxidized, passivated Ti upon exposure to ambient air) protects the underlying layers from oxidation.  

The FeV-Ni/Nb series has the same stack structure as FeV/Nb, except that 4-nm-thick Ni is inserted between FeV and Nb. The Ni thickness of 4 nm was chosen to be greater than the spin decay length of $\approx$2-3 nm [Refs.~\citen{lim2022absorption, Ko2020optical}] and less than the ferromagnetic exchange length of $\approx$8 nm [Ref.~\citen{abo2013definition}]. It is  also sufficiently thick for the layer-resolved x-ray magnetic circurlar dichroism (XMCD) measurement, detailed below, but still much thinner than the 20-nm-thick FeV layer that dominates the large-signal FMR dynamics.
%In principle, a single layer of Ni could be used as the pumping source; in reality, its high intrinsic damping parameter of $\sim$ 0.03 [Ref.~\citen{schoen2017magnetic}] leads to a poor signal-to-noise ratio in broadband FMR, making it difficult to resolve an increase in damping above the noise floor. In the FeV(20)-Ni(4) bilayer source, the Ni magnetization is exchange-coupled to the magnetization of FeV, which has an order of magnitude lower intrinsic damping parameter of $< 0.003$ [Refs.~\citen{smith2020magnetic} and \citen{arora2021magnetic}]. The Ni magnetization coupled to \textcolor{blue}{the} resonantly driven FeV yields an adequately large signal-to-noise ratio for broadband FMR spectra. 

\begin{figure}[bt]
\centering
  \includegraphics[width=0.90\columnwidth]{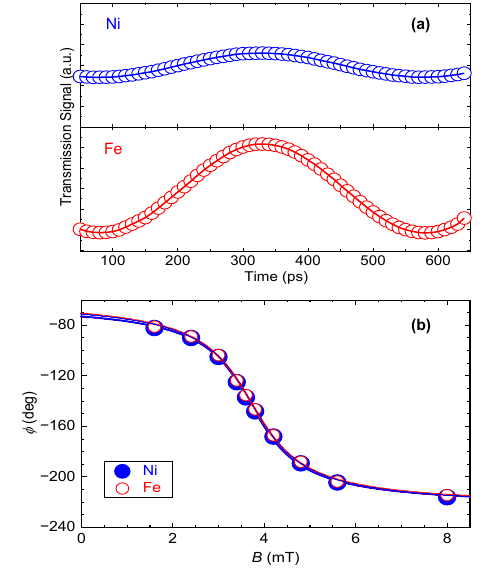}
\caption{(a) Time traces of XFMR signals acquired at 2 GHz under an in-plane applied field of 3.7 mT and (b) field dependence of the precessional phase $\phi$ at the Ni and Fe $L_3$ edges in FeV-Ni (\tNb = 0).}
\label{fig:phase}
\end{figure}

%%% Verification of FeV-Ni Coupling %%%
Verification of the coupling between FeV and Ni under FMR requires a layer-resolved measurement. To this end, we separately acquired time traces of the precessing magnetizations of FeV and Ni through XMCD at Beamline 4.0.2 at the Advanced Light Source. The details of this dynamic XMCD method, also known as x-ray FMR or XFMR, are found in Refs.~\citen{klewe2023observation} and  \citen{klewe2020element}. In brief, we conducted dynamic XMCD measurements on an FeV-Ni sample with \tNb = 0 nm, grown on Ce-doped Y$_3$Al$_5$O$_{12}$ (YAG) substrates for luminescence yield detection of XMCD~\cite{piamonteze2020soft}. The signals at the Fe $L_3$ edge (from FeV) and the Ni $L_3$ edge were measured under an in-plane applied quasistatic field and a microwave excitation of 2 GHz. The time traces in Fig.~\ref{fig:phase}(a), acquired at the resonance field (\Bres\ $\approx 3.7$ mT), reveal sinusoidal oscillations of the excited Fe and Ni moments with a period of 500 ps. Quick visual inspection suggests that the Fe and Ni precess in phase, consistent with the coherent coupling of the magnetizations in FeV and Ni. 
We fit the $B$ dependence of the precessional phase $\phi$ [Fig.~\ref{fig:phase}(b)] with $\phi = \phi_0 + \arctan(\Delta B/(B-$\Bres$))$. The baseline phases $\phi_0$ for Fe and Ni are within
$\approx$$1^\circ$ of each other, which can be accounted for by the 3-ps timing jitter of the master oscillator of the beamline. The deviation between Fe and Ni for the obtained FMR linewidth $\Delta B$ and resonance field \Bres\ are also within the field reading uncertainty of $\pm$0.2 mT at the beamline. 

Thus, our quantitative dynamic XMCD results confirm rigid coupling between resonantly excited magnetizations of the FeV and Ni layers. This is reasonable considering the direct interface between the FeV and Ni layers, along with the 4-nm Ni layer thickness under the ferromagnetic exchange length of $\approx$8 nm [Ref.~\citen{abo2013definition}]. While this measurement was performed in the in-plane geometry due to experimental constraints, we expect this strong ferromagnetic exchange to enforce coherent precession in the out-of-plane FMR geometry as well. This is because the ferromagnetic exchange interaction, which is the primary mechanism for this coupling, is isotropic and does not depend on the direction of the applied field.
%ensuring uniform dynamics across the Ni thickness.  

Spin/orbital pumping occurs across the interface between the source and the sink. Therefore, in FeV-Ni/Nb, pumping into Nb must originate from the magnetization dynamics of Ni in direct contact with Nb, rather than the dynamics of FeV away from Nb. Any spin current pumped from FeV largely decays within the first $\approx$2-3 nm of Ni~\cite{lim2022absorption, Ko2020optical}, rather than propagating through the 4-nm-thick Ni layer. Orbital transport from FeV through Ni is less clear, but orbital pumping by FeV is plausibly just a few percent relative to spin pumping~\cite{go2025orbital} and can be neglected. 

%%% Out-of-Plane FMR %%%
We evaluate our hypothesis that FeV-Ni/Nb could display substantial orbital pumping from Ni to Nb, distinct from FeV/Nb with limited orbital pumping. For this purpose, we conducted broadband FMR spectroscopy, using a flip-chip coplanar-waveguide setup with the quasistatic field applied \emph{out of the film plane}. Crucially, this out-of-plane configuration eliminates extrinsic non-Gilbert two-magnon relaxation~\cite{hurben1998theory}, thereby enabling reliable quantification of the Gilbert damping parameter~\cite{boone2015spin,boone2013spin} emerging from spin/orbital pumping. Our electromagnet with a maximum field of 2 T is sufficient to saturate FeV(-Ni), with a demagnetizing field of $\mu_0 M_\mathrm{s} \approx$1.1 T, completely out of plane. Each FMR spectrum is fit with a Lorentzian derivative to extract the resonance field \Bres\ and the half-width-at-half-maximum FMR linewidth \LW\ [Refs.~\citen{smith2020magnetic, lim2022absorption}].

\begin{figure*}[htb]
        \centering
        \includegraphics[width=0.95\textwidth]{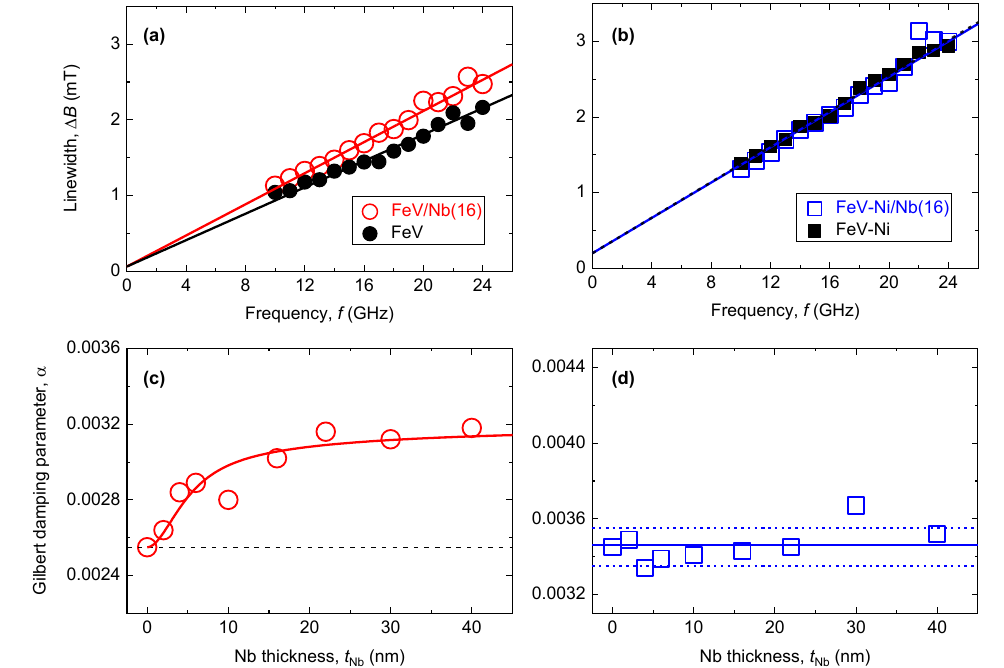} % Adjust width as needed
        \caption{(a,b) Half-width-at-half-maximum FMR linewidth $\Delta B$ as a function of frequency $f$ for (a) FeV/Nb and (b) FeV-Ni/Nb, measured with the samples magnetized out-of-plane. The linear fits are used to quantify the Gilbert damping parameter $\alpha$ with Eq.~\ref{eq:LW}. (c,d) Gilbert damping parameter $\alpha$ plotted against the Nb thickness for (c) FeV/Nb and (d) FeV-Ni/Nb. The horizontal dashed line in (c) denotes $\alpha_\mathrm{no-sink}$ for FeV (\tNb = 0). In (d), the solid line denotes the average value of $\alpha$ over all FeV-Ni/Nb samples, whereas the dashed lines denote the standard deviation.}
        \label{fig:LW_OP_damping}
    \end{figure*}

%A more critical insight is gained from the linewidth $\Delta B$. In particular, 
The slope of $\Delta B$ vs $f$ [Fig.~\ref{fig:LW_OP_damping}(a,c)] permits straightforward quantification of the Gilbert damping parameter $\alpha$ via
    \begin{equation}
      \Delta B = \Delta B_{\text{0}} + \frac {h}{g\mu_\mathrm{B}} \alpha f,
      \label{eq:LW} 
    \end{equation}
where $h$ is Planck's constant, $g \approx 2.1$ is the $g$-factor (derived from the $f$ dependence of \Bres\ as shown in the Supplementary Material), and $\mu_\mathrm{B}$ is the Bohr magneton. 
The zero frequency linewidth $\Delta B_\mathrm{0}$, capturing the broadening of the FMR linewidth from magnetic inhomogeneity~\cite{mewes2015relaxation}, is $\Delta B_\mathrm{0} = 0.06$  mT for FeV/Nb and 0.2 mT for FeV-Ni/Nb.
%accounts for only a small fraction of the total $\Delta B$ -- i.e., $\Delta B_\mathrm{0} = 0.06$  mT for FeV/Nb and 0.2 mT for FeV-Ni/Nb.
%such that $\alpha$ is the sole free parameter in the fit. We also obtain a similar outcome by fitting with $\Delta B_\mathrm{0}$ as an additional free parameter, as detailed in the Supplementary Material. 
We apply Eq.~\ref{eq:LW} to data at $f \geq 10$ GHz to quantify $\alpha$. Below 10 GHz (or $B \lesssim 1.4$ T), $\Delta B$ exhibits a slightly shallower slope with decreasing $f$ due to incomplete saturation; our analysis in the Supplementary Material presents fits accounting for the nonlinearity of these low-frequency data points, which yield $\alpha$ nearly identical to the values in Fig.~\ref{fig:LW_OP_damping}(c,d).  
%As an additional check, we also fit $\Delta B$ over the entire $f$ range by incorporating a nonlinear phenomenological term scaling as 1/$f^n$~\cite{Hsueh2011} into Eq.~\ref{eq:LW}. This alternative protocol and the simple linear fit again yield similar outcomes [see the Supplementary Material]. 

As shown in Fig.~\ref{fig:LW_OP_damping}(a), FeV interfaced with Nb exhibits a noticeably steeper slope in $\Delta B$ vs $f$, i.e., greater Gilbert damping, compared to FeV without Nb. 
With increasing \tNb, $\alpha$ increases from $\approx$0.0026 at \tNb = 0 to a saturated level of $\approx$0.0032 [Fig.~\ref{fig:LW_OP_damping}(c)]. 
This trend is consistent with the pumping of angular momentum from the FeV source to the adjacent Nb sink in which the angular momentum decays [Fig.~\ref{fig:cartoons}(a)]. 

Further insights can be gained by fitting the \tNb\ dependence of $\alpha$ with a diffusion model. We remark that no established theoretical model exists yet for orbital diffusion. Nevertheless, since spin pumping likely overwhelmingly dominates over orbital pumping from an Fe-based source~\cite{go2025orbital}, we analyze the \tNb\ thickness dependence of $\alpha$ in FeV/Nb with the spin diffusion model~\cite{haney2013current, boone2015spin}, 
\begin{equation}
    \alpha = \alpha_\text{no-sink} + \frac{g \mu_B \hbar}{2 e^2 M_\text{s} t_\text{M}} \left[ \frac{1}{G_{\uparrow \downarrow}} + 2 \rho_{\text{Nb}} \lambda_\text{d} \coth \left( \frac{t_{\text{Nb}}}{\lambda_\text{d}} \right) \right]^{-1},
    \label{eq:diffusion}
\end{equation}
with the saturation magnetization $M_\text{s} = 900$ kA/m and the magnetic source thickness $t_\text{M} = 20$ nm. The thickness dependent resistivity of Nb is modeled as $\rho_\mathrm{Nb} = 1.5\times10^{-7} ~\Omega\text{m}+ 1.4\times10^{-15}~\Omega\text{m}^2/t_\mathrm{Nb}$ [Ref.~\citen{jeon2018enhanced}].  The free parameters of the fit here are the spin-mixing conductance $G_{\uparrow \downarrow}$ and the spin diffusion length $\lambda_\text{d}$. We obtain $G_{\uparrow \downarrow} = (5.8\pm2.1)\times10^{14}$ $\Omega^{-1}$m$^{-2}$ ($g_{\uparrow \downarrow} = (h/e^2)G_{\uparrow \downarrow} = 15\pm5$ nm$^{-2}$), in line with typical values of ferromagnet/nonmagnetic-metal interfaces~\cite{zhu2019effective,zwierzycki2005first}; $\lambda_\text{d} = 5.5\pm1.0$ nm 
%is much shorter than $> 30$ nm estimated in Refs.~\citen{jeon2018enhanced} and \citen{jeon2018spin}
is similar to $\approx 3$ nm [Ref.~\citen{liu2023giant}] and $\approx 8$ nm [Ref.~\citen{boone2013spin}] reported by others. 

We also evaluate the \emph{effective} spin-mixing conductance $G^{\uparrow \downarrow}_\mathrm{eff}$, which is directly proportional to the saturated Gilbert damping enhancement $\Delta \alpha$, as given by~\cite{tserkovnyak2005nonlocal, zhu2019effective} 
\begin{equation}
    G^{\uparrow \downarrow}_\mathrm{eff} = \frac{2e^2 M_\mathrm{s} t_\mathrm{M}}{g\mu_\mathrm{B}\hbar}\Delta \alpha. 
    \label{eq:Geff}
\end{equation}
FeV/Nb exhibits $\Delta \alpha \approx 0.0006$, corresponding to an effective spin-mixing conductance of $G^{\uparrow \downarrow}_\mathrm{eff} \approx 3 \times10^{14} $ $\Omega^{-1}$m$^{-2}$ ($g^{\uparrow \downarrow}_\mathrm{eff}  \approx 7$ nm$^{-2}$). These values are again in line with reasonable values for ferromagnet/normal-metal interfaces~\cite{zhu2019effective}. Overall, the conventional framework of spin pumping (e.g., Eqs.~\ref{eq:diffusion} and \ref{eq:Geff}) reasonably captures our experimental results for FeV/Nb. 

The observed trend for FeV-Ni/Nb is qualitatively different, as seen in Fig.~\ref{fig:LW_OP_damping}(b,d), which constitutes the key results of our work. FeV-Ni with and without a Nb layer exhibit nearly identical slopes in $\Delta B$ vs $f$  [Fig.~\ref{fig:LW_OP_damping}(b)]. As summarized in Fig.~\ref{fig:LW_OP_damping}(d), the Gilbert damping parameter of FeV-Ni/Nb remains essentially constant with \tNb: FeV-Ni with no Nb exhibits $\alpha_\mathrm{no-sink} \approx 0.0035$, and $\alpha$ remains at that value within the scatter of $\pm 0.0001$ for all FeV-Ni/Nb samples. This lack of detectable enhancement of $\alpha$ with the addition of Nb indicates no significant pumping from Ni to Nb. 

As confirmed by our dynamic XMCD experiment [Fig.~\ref{fig:phase}], FeV and Ni magnetizations precess together coherently. Therefore, Ni is not just a passive sink of angular momentum, but also an active source that can pump angular momentum into adjacent layers. The constant damping parameter of $\alpha = 0.0035\pm0.0001$ for FeV-Ni/Nb, about 35\% higher than that of FeV, may be due to mutual angular momentum pumping between ferromagnetic FeV and Ni~\cite{li2020coherent}. The origin of this \tNb-independent elevated damping in FeV-Ni remains the subject of a future study.

The key finding here is that spin/orbital pumping from Ni to Nb is so small that it is below our present detection limit. We take the scatter in $\alpha$ of 0.0001 to be the upper limit of $\Delta \alpha$ for FeV-Ni/Nb. With $t_\mathrm{M} = 24$ nm and $M_\mathrm{s} = 870$ kA/m in Eq.~\ref{eq:Geff}, the effective mixing conductance at the Ni/Nb interface would be $G^{\uparrow \downarrow}_\mathrm{eff} < 0.5 \times10^{14} $ $\Omega^{-1}$m$^{-2}$ ($g^{\uparrow \downarrow}_\mathrm{eff} < 1$ nm$^{-2}$). This estimated angular-momentum transmission across Ni/Nb is an order of magnitude weaker than transmission across FeV/Nb. 
%In semi-quantitative terms, the spin-mixing conductance (possibly entangled with orbital-mixing conductance) for the Ni/Nb interface is $G_{\uparrow\downarrow}\lesssim 1\times10^{14}$ $\Omega^{-1}$m$^{-2}$ (or $g_{\uparrow\downarrow}\lesssim 3$ nm$^{-2}$), at least several times smaller than that for the FeV/Nb interface.   

%%% Commentary/Implications %%%
The absence of resolvable spin and orbital pumping from Ni to Nb, in contrast to pronounced pumping from FeV to Nb, is a striking finding. It highlights that the composition of the ferromagnetic source can have a dramatic impact on the efficiency of angular-momentum transfer into an adjacent metal. This finding is consistent with early computational predictions that the matching of electronic band states at the Fermi energy between the ferromagnetic and nonmagnetic metals 
greatly impacts the strength of spin pumping~\cite{zwierzycki2005first}.
It is possible that a significant band mismatch exists at the Ni/Nb interface and suppresses angular-momentum transfer in our samples. However, the extent of the band mismatch at this interface, along with its precise correlation with angular-momentum pumping (or the lack thereof), remains an open question.
Further theoretical work is warranted to identify the exact mechanism responsible for the suppression of significant pumping. 

Our findings underscore the potential complexity of prior studies that aim to detect angular-momentum transport electrically from lateral voltage signals, which can have coexisting contributions from both the ferromagnetic and nonmagnetic metals~\cite{davidson2020perspectives,go2020theory,kim2024spin}. While our Gilbert damping measurements show minimal angular-momentum pumping from Ni to Nb, prior electrical measurements may be influenced by other coexisting phenomena.
%In light of our findings, past reports of robust orbital transport between Ni and Nb ~\cite{dutta2022observation, bose2023detection, liu2023giant} (and perhaps other similar transition metals such as Ti~\cite{hayashi2023observation,choi2023observation,hayashi2024observation}) may need to be re-evaluated. This may apply to both orbital pumping and its reciprocal effect, orbital torques. Previous pumping and torque experiments often used lateral electrical voltage measurements, which can combine signals from both the ferromagnetic metal and the nonmagnetic metal~\cite{hayashi2024observation,davidson2020perspectives,go2020theory,kim2024spin,lee2021orbital,sala2022giant}. 
For example, electrically detected pumping can include signals like rectification, thermoelectric effects, and inverse spin-Hall signals originating from within the ferromagnet itself~\cite{bai2013universal,schultheiss2012thermoelectric,tsukahara2014self}, potentially augmented by asymmetric interfaces~\cite{chen2016robust, kim2020generalized}. Similarly, electrically detected torque measurements can be influenced by ``self-torques'' generated inside the ferromagnet with asymmetric interfacial and bulk properties~\cite{Wang2019anomalous,kim2020generalized, maizel2024vertically}. In general, disentangling angular-momentum transfer between the two materials from other coexisting effects is challenging with these voltage detection methods. 

Our method here monitors angular-momentum dissipation through Gilbert damping in the out-of-plane FMR configuration. It does not rely on lateral voltage measurements, so it is free of complications that hamper electrically detected pumping and torque measurements. Our approach may provide a complementary -- and potentially more reliable -- view of angular-momentum transport in magnetic heterostructures. While we did not find significant orbital pumping in Ni/Nb, we do not universally rule out the existence of orbital pumping in other systems. Critical insights into orbital pumping may be gained through future studies that quantify Gilbert damping enhancement with high precision, particularly in heterostructures with ferromagnetic sources possessing different spin/orbital characteristics.

%Some studies have bypassed the above-discussed complications by employing insulating magnetic materials~\cite{ding2020harnessing,santos2023inverse,wang2025orbital}. For example, a recent experiment reveals distinct electrically detected pumping \textcolor{blue}{signals} with insulating YIG and Bi:YIG magnetic sources, which can be plausibly explained by Bi:YIG generating comparatively strong orbital pumping than YIG~\cite{wang2025orbital}. If orbital pumping emerges from insulating magnets, it appears reasonable that ferromagnetic metals should also exhibit some orbital pumping. While not detected in Ni/Nb in our present study, orbital pumping may be resolved in other metallic heterostructures through future studies that quantify Gilbert damping enhancement with high precision. 

In summary, we have attempted to find a signature of orbital pumping from Ni to Nb. Specifically, we have compared FMR results of heterostructures without Ni (FeV/Nb) and with Ni (FeV-Ni/Nb). The FeV and Ni magnetizations in FeV-Ni are confirmed to be coupled coherently, boosting the precession amplitude of Ni for spin/orbital pumping. Out-of-plane broadband FMR measurements allow for straightforward quantification of Gilbert damping. We observe an increase and saturation of damping with Nb thickness in FeV/Nb, which is consistent with spin pumping from FeV to Nb. However, no detectable change in damping is observed for FeV-Ni/Nb. This surprising observation implies minimal spin and orbital pumping from Ni to Nb. The absence of detectable pumping here suggests the need to reconsider angular momentum transport in heterostructures consisting of Ni interfaced with Nb (and perhaps other similar transition metals). While our study does not rule out orbital pumping in general, it highlights the current challenge in uncovering this phenomenon in metallic multilayers.

\textbf{Supplementary Material: } The Supplementary Material presents the quantification of the $g$-factor and different protocols of fitting the frequency dependence of the out-of-plane FMR linewidth.

\begin{acknowledgments}
O.A.B., R.E.M, and S.E. were supported by the National Science Foundation (NSF) under Grant No. ECCS-2144333. G.T.S. and S.A.K. were supported by the NSF under Grant No. ECCS-2236160. This research used resources of the Advanced Light Source, a U.S. DOE Office of Science User Facility under Contract No. DE-AC02-05CH11231. SE also thanks support by the Luther and Alice Hamlett Junior Faculty Fellowship. 
\end{acknowledgments}

\section*{Data Availability}
The data that support the findings of this study are available from the corresponding authors upon reasonable request.

%\bibliography{Refs_with_DOI}


\end{document}